\documentstyle{mn}
\include{psfig}
\begin{document}
\newcommand{\Ha}{H$\alpha$}
\newcommand{\Ham}{H\alpha}
\newcommand{\Hbm}{H\beta}
\newcommand{\Hb}{H$\beta$}
\newcommand{\oiii}{{\rm \mbox{[O \sc iii]}\ }}
\newcommand{\oii}{{\rm \mbox{[O \sc ii]}\ }}
\newcommand{\nii}{{\rm \mbox{[N \sc ii]}\ }}
\newcommand{\sii}{{\rm \mbox{[S \sc ii]}\ }}
\newcommand{\niioiii}{{\rm \mbox{[N \sc ii]}}/{{\rm \mbox{[O \sc iii]}}\ }}
\newcommand{\oiiinii}{{\rm \mbox{[O \sc iii]}}/{{\rm \mbox{[N \sc ii]}}\ }}
\newcommand{\niioii}{{\rm \mbox{[N \sc ii]}}/{{\rm \mbox {[O \sc ii]}}\ }}
\newcommand{\siioii}{{\rm \mbox{[S \sc ii]}}/{{\rm \mbox {[O \sc ii]}}\ }}
\newcommand{\ohb}{{\rm \mbox{[O \sc iii]/{H$\beta$}}\ }}
\newcommand{\oiihb}{{\rm \mbox{[O \sc ii]/{H$\beta$}}\ }}
\newcommand{\oiioiii}{{\rm ([O \sc ii] + [O \sc iii])/{H$\beta$}\ }}
\newcommand{\OiiOiii}{{\rm [O \sc ii]/[O \sc iii]\ }}
\newcommand{\hahb}{H$\alpha$/H$\beta$}
\newcommand{\nha}{{\rm \mbox{[N \sc ii]/H$\alpha$}\ }}
\newcommand{\sha}{{\rm \mbox{[S \sc ii]/H$\alpha$}\ }}
\newcommand{\hii}{{H \sc ii}\ }
\newcommand{\HII}{{H \sc ii}\ }
\newcommand{\hi}{{H \sc i}\ }
\newcommand{\om}{{\rm \mbox{[O \sc iii]}}}
\newcommand{\nm}{{\rm \mbox{[N \sc ii]}}}
\newcommand{\kms}{\rm km s$^{-1}$}

\title[O/H in NGC 1313]
{The O/H Distribution in the Transition Magellanic Galaxy NGC 1313}

\author[Walsh \& Roy]
{J. R. Walsh$^1$ and J.-R. Roy$^2$\\
$^1$ European Southern Observatory, Karl-Schwarzschild-Strasse 2,
D-85748 Garching bei M\"unchen, Germany\\
$^2$ D\'epartement de physique and Observatoire du mont M\'egantic,
Universit\'e Laval, Qu\'ebec Qc G1K 7P4, Canada}

\maketitle
\begin{abstract}
Multi-fibre emission-line spectrophotometry of 33 \hii regions
and 3 diffuse interstellar medium positions are presented for the barred
Magellanic galaxy NGC 1313. The \hii regions show a fairly narrow
range of thermal conditions characteristic of high excitation
nebular gas. Electron temperature was directly determined in four
of the \hii regions.
The global O/H abundance distribution appears very flat across the disk
at 12 + log O/H $\approx$ 8.4 $\pm$ 0.1,
with possibly  the bar regions having abundances higher 
by 0.2 dex than the outer disc. NGC 1313 is the  highest mass barred galaxy 
known {\it not} to have any radial abundance gradient. The key role of the
bar on the abundance distribution in disc galaxies is revisited.
\end{abstract}

\section{Introduction}

NGC 1313 has been described by de Vaucouleurs \shortcite{dV63} as a galaxy in transition
between SBm and SBc type. It is sufficiently close to the
Magellanics to recall some of their properties, and at the same
time it can be compared to M33 or NGC 300.
Low-mass irregulars like the Magellanic Clouds \cite{PA78}, NGC 2366
and NGC 4395 \cite{RO96A} and NGC 4214 \cite{KS96} have very
flat abundance distributions, with mean metallicity levels which appear to scale
with the overall mass of the galaxies. Edmunds \& Roy \shortcite{ER93}
have noted that abundance gradients
seem to disappear at the same absolute magnitude ($M_{\rm B}$ $\sim$ -17
in late-type spirals) where spiral structure no longer exists. In contrast,
late-type spirals of Sc type, like NGC 300, NGC 7793 and M 33, which
are slightly more massive than the above low mass irregulars, show well developed 
global O/H abundance gradients of the order of --0.08$\pm$0.02 dex/kpc.

 It is also well established that, compared to normal disc galaxies,
 barred galaxies have flatter radial abundance
distributions whatever their mass (Vila-Costas \& Edmunds \shortcite{VE92};
Zaritsky et al. \shortcite{ZA94}; Martin \& Roy \shortcite{MR94}; Roy
\shortcite{RO96B}).  Also well-known are the barred structures of many
Magellanic irregulars. A bar can be  reponsible for angular momentum transfer
by inward and outward radial flows across the disc; the
bar becomes then  an effective way for radial
homogenization of the abundance distribution \cite{FR94}. Thus the presence
of strong bar features in SBm type galaxies may not be foreign to
the uniform abundance distribution, as originally suggested
by Pagel et al. \shortcite{PA78} \shortcite{PA80}.
Nevertheless the difference in behaviour
of the chemical composition between  galaxies of the same morphological
type, but of different bar strength, is intriguing. 

It would also be helpful to establish at what morphological type,
what luminosity and what mass abundance gradients become established in
galaxy disks.  Somewhere between units of M33 size and mass,
and units of Large Magellanic Cloud (LMC) size, the discs evolve in ways which enable
the buildup of  
an abundance gradient. Because of its bar and its intermediary properties, NGC 1313 is
an attractive object to study the ``take-off'' of abundance gradients in
galaxy discs.

In order to improve the understanding of the origin and
evolution of abundance gradients, we have obtained fibre spectrophotometry of 
33 \hii regions in the late-type galaxy NGC 1313. Section 2 provides some details of the
galaxy and on how the \hii regions were selected; section 3 presents the results and in
section 4 the features of the O/H gradient and comparison with other galaxies of
similar luminosity are discussed.    

\section{Observations and data reduction}

\subsection{The galaxy NGC 1313}

NGC 1313 is an isolated very late-type barred spiral which
has long been thought to have a complicated history.
Its morphology is reminiscent of several of the
irregular  or amorphous objects seen in the Hubble Deep Field. Its
ill-formed arms have led to speculation about the
galaxy being the result of a tidal interaction (Peters et al. \shortcite{PE94}).
However, from their
recently completed thorough optical and \hi study,
Ryder et al. \shortcite{RY95} concluded that NGC 1313 is dynamically
and morphologically more regular and well developed than the
class of Magellanic barred spirals. Bright \hii regions outline
the bar and  the arms, with the most luminous being in the arms,
particularly the northwestern one \cite{MG83}. There
is no colour gradient across the galaxy \cite{RY95}. 
A Type II supernova was observed in NGC 1313 in 1968 \cite{SE68};
another Type II supernova (SN1978K) was found retrospectively
by Ryder et al. \shortcite{RY93B}. Table 1
gives some of the general properties of the galaxy.

\begin{table}
\caption{Global properties of NGC 1313} 
\begin{tabular}{lc}
\hline \hline
Parameter & Value \\
\hline  
$\alpha$ (J2000) & 3$^{\rm h}$18$^{\rm m}$15.$^{\rm s}$5 \\
$\delta$ (J2000) &  -66$^\circ$29$'$51$''$ \\
Morphological type & SB(s)d\\
Inclination & 48$^\circ$\\
Position angle & 0$^\circ$ \\
Galactic extinction ($A_{\rm B})$ & 0.04\\
Systemic velocity (km s$^{-1})$ & 480 \\
$\rho_0$  & 4.$'$56\\
$b/a(i)$ & 0.63 \\
D (Mpc) & 4.5 \\
Scale (pc arcsec$^{-1}$) & 22 \\
\hline
\end{tabular}\\
Notes: (a) Position, angles and velocity are from Ryder et al. 
\shortcite{RY95}; (b)
 type, extinction and isophotal radius are from de Vaucouleurs et al. 
\shortcite{dV91}; (c) stellar bar axis ratio, $b/a(i)$, corrected for
inclination is from Martin \shortcite{MA95}; (d) distance is from de Vaucouleurs 
\shortcite{dV63}.
\end{table}

Pagel et al. \shortcite{PA80} presented the results of the spectrophotometry
of six \hii regions in NGC 1313  (and seven in NGC 6822)
and concluded that the abundance
distribution was quite uniform at 12 + log O/H = 8.26 $\pm$ 0.07.
From spectrophotometric measurements of 14 \hii regions,
Ryder \shortcite{RY93A}
derived a steep abundance gradient of --0.09 dex kpc$^{-1}$, but
this result depends on two  points at large effective radius which strongly
affected the derived slope.
 
\begin{figure*}
\centerline{\psfig{file=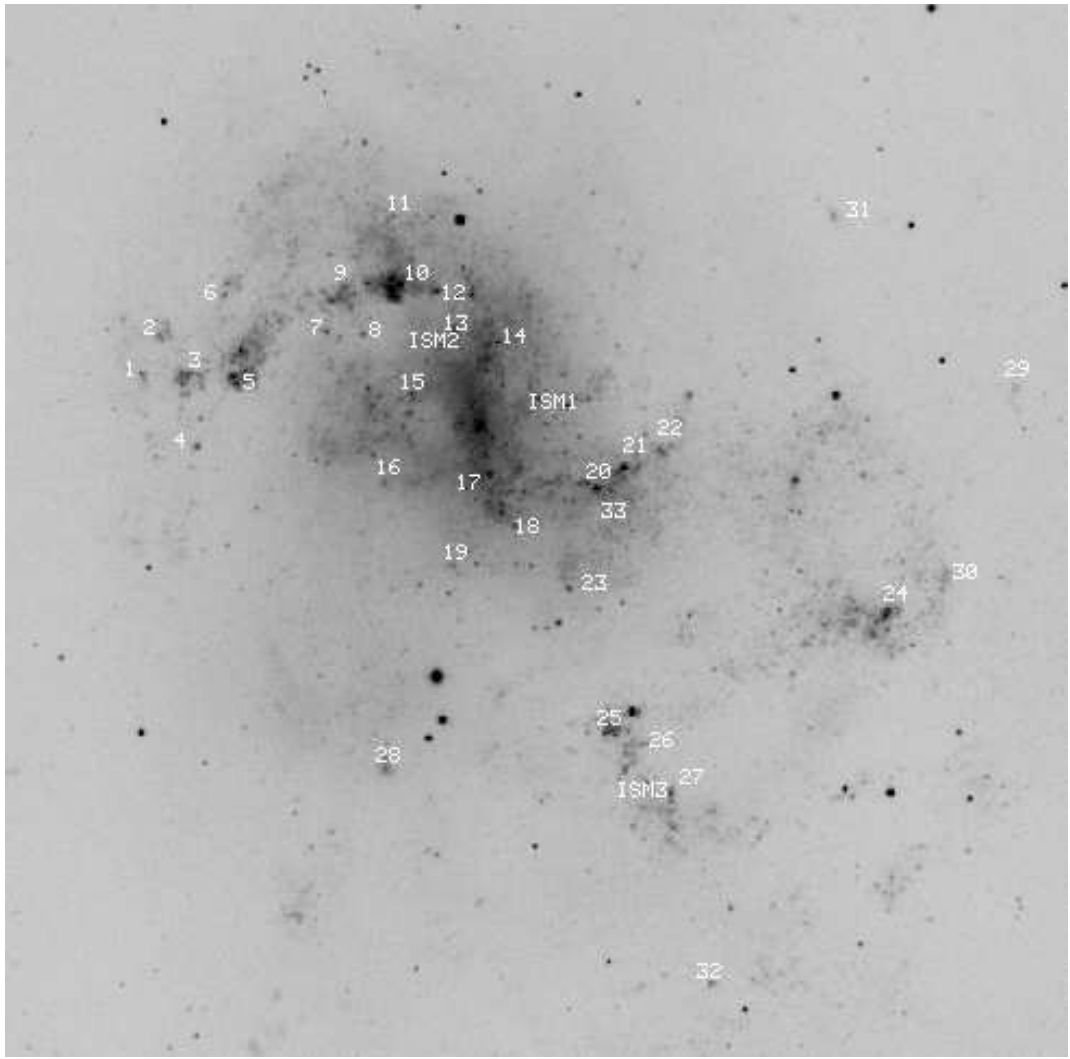,height=14.0cm,clip=}}
\caption{A B-plate of NGC 1313, obtained by M. Marcelin with the ESO 3.6-m telescope,
is shown. The numbers correspond to the \hii regions for which  spectra were 
obtained; ISM{\em n} corresponds to positions of spectra of the diffuse ionized 
emission.}
\label{fig1}
\end{figure*}

\subsection{Selection of H~{\sc II} regions}

Marcelin \& Gondoin \shortcite{MG83} catalogued 375 \hii regions in
NGC~1313 from analysis of an H$\alpha$ plate. Michel Marcelin kindly loaned a B
plate of 40 min exposure (taken with the ESO 3.6-m at Cassegrain focus on IIaO with a GG385 
filter centred at 6575\AA\ on 1978 January 31) which was scanned 
on the ESO PDS machine. The scanning aperture was 50$\mu$m square and the sampling
25$\mu$m (0.474$''$). The image is shown in Figure 1. Using the catalogue of
Marcelin \& Gondoin \shortcite{MG83}, 104 \hii regions were identified
which were suitable for fibre spectroscopy in that they were not too faint,
distinct and with a representative distribution over the bar and
arms. The centroid positions of these \hii regions were
determined and also of
12 HST GSC stars by 2-D Gaussian fitting. The RA and Dec of the \hii regions were then
determined by a six coefficient astrometric solution. The projected size of 
NGC~1313 is such that all the ($\sim$50) fibres in a single FOCAP bundle could not be  filled by \hii regions since they
are too close (the closest distance between fibres is 19$''$); for this
reason we could not observe as many regions in the bar as we would have liked.
A total of
37 \hii regions were observed with fibres. A number of
positions were also identified to sample the diffuse interstellar medium of the galaxy
and examine their emission line spectrum; they are denoted by ISM1 - ISM3 on Figure 1.
For the purposes of sky subtraction, ten positions away from the galaxy
and from stars were identified for sky fibres. Of the 37 \hii regions observed,
33 had strong enough emission (i.e. at least H$\alpha$ and H$\beta$) for detailed
analysis.  Three of the ISM regions had spectra strong enough for individual
analysis (7 were observed). The observed \hii regions and ISM positions are identified
on Figure 1 by a running number and an ISM tag respectively.

\subsection{Observations}
Low dispersion spectra of the selected \hii regions in NGC 1313 were
obtained on the night of 1993 December 12-13 with the
Anglo-Australian Telescope, using the fiber optics fed FOCAP system,
the RGO spectrograph, and the Tek \#2 CCD.
The same instrumental set-up with 300$\mu$m (2.0$''$) fibres was employed as for 
the observations of NGC 1365 reported in Roy \& Walsh \shortcite{RW96}. The total
exposure time was 6200s, divided mostly into 1000s exposures. Seeing was about
1$''$ but there was some cloud encountered during the observations. 37 \hii regions,
10 sky positions and 7 
ISM positions off identified \hii regions were observed.
Full details of the observation procedures and 
a detailed discussion of the observation and reduction steps followed with the 
AAT FOCAP fiber optics system can be found in Roy \& Walsh \shortcite{RW96}.

\subsection{Reduction of the spectra}

The data were reduced identically to the NGC 1365 \hii region fibre data
\cite{RW96}. The average spectrum over three nights of the spectrophotometric
standard L870-2 \cite{OK74} was employed for flux calibration. Gaussians were
fitted to the emission lines and polynomials to the underlying continuum
using an interactive procedure and line fluxes and photon noise errors on the
line fluxes were expressed relative to $F(H\beta)$ = 100. The magnitude of
interstellar reddening was determined from the H$\alpha$/\Hb \ ratios;
comparison was done to the
theoretical decrement as given by Brocklehurst (1971) for a temperature
of 10,000 K and a density of 100 cm$^{-3}$, after adding 2 \AA\
of equivalent width to the \Hb \ emission line to compensate for
the underlying Balmer absorption (cf. McCall, Rybski \& Shields 1985;
Roy \& Walsh 1987). The observed spectra were corrected for extinction
as a function of wavelength using the standard
reddening law of Seaton \shortcite{SE79},
as specified by Howarth \shortcite{HO83} and using R=3.1.

\section{Results}

\begin{figure*}
\centerline{\psfig{file=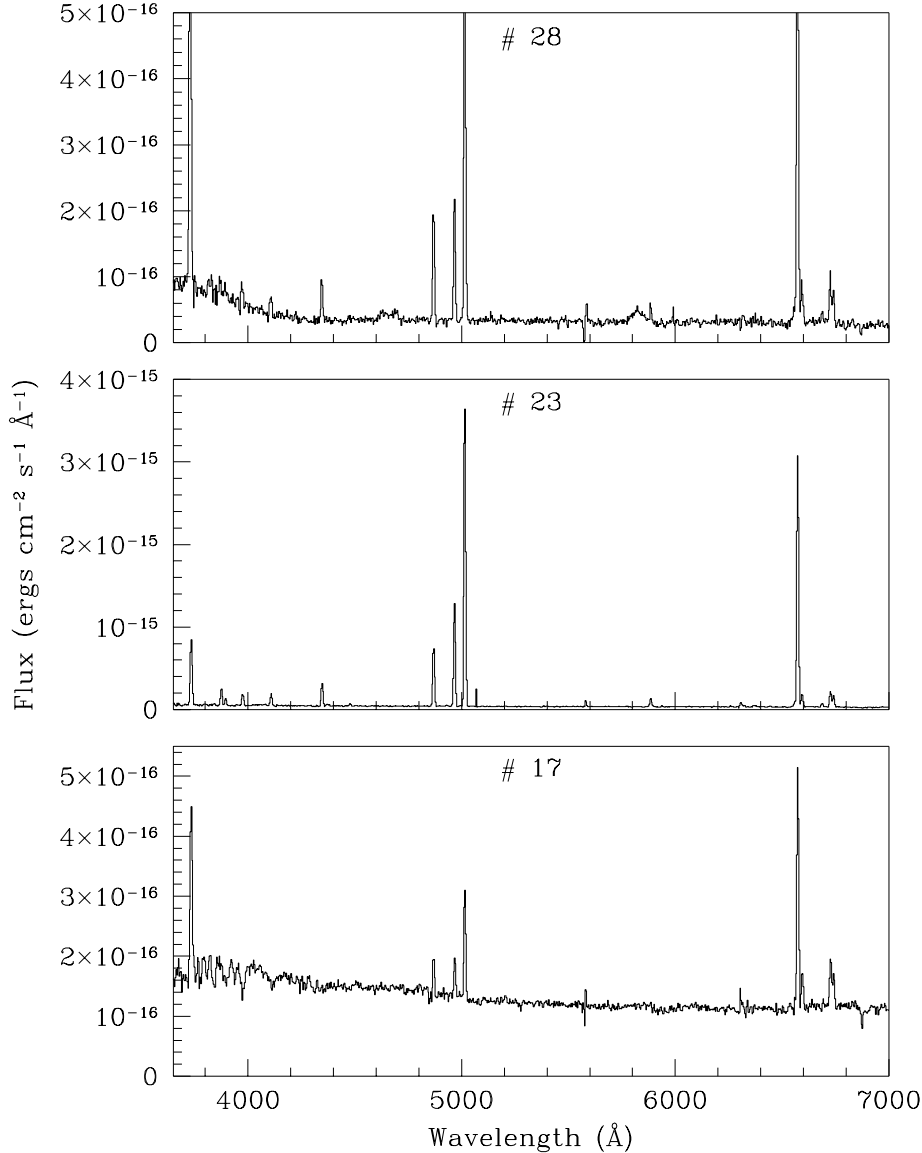,height=16.0cm,clip=}}
\caption{Examples of spectra of \hii regions at different galactocentric 
distances in NGC 1313. Exact positions of the fibre positions and details
of the spectra are listed in Table 2. 
The strong signature of Wolf-Rayet stars in region \#28 is noteworthy.}
\label{fig2}
\end{figure*}

Figure 2 shows three examples of some of the  best spectra; all those
retained were characterized by high signal-to-noise, except
for those three corresponding to positions on the diffuse ionized
interstellar medium. Signatures of Wolf-Rayet stars were found in
regions \#3 and \#28, both at large galactocentric radii. The spectra 
are in general representative of high excitation \hii regions, and that level is
rather uniform across the disk. The strength of
the \oiii 4363\AA\ line was strong enough to determine
the electron temperature in 4 \hii regions; they
are regions \#23 (T$_e$ = 12,250$\pm$400 K), \#5 (T$_e$ =
11,500$\pm$500 K), \#21 (T$_e$ = 11,000$\pm$700 K)  and \#12
(T$_e$ = 9,900$\pm$600 K). Extinction is
moderately high and uniform across the disc at $c$(\Hb) $\sim$
0.7. Positions of the \hii regions (in terms of RA and Dec offset from
the galaxy centre specified in Table 1) and reddening-corrected line fluxes 
relative to F(H$\beta$)=100 are given in Table 2. The final column lists the
deprojected fractional radius in terms of the isophotal radius, $\rho_{0}$,
listed in Table 1. 

Three of the observed \hii regions were also studied by Pagel et
al \shortcite{PA80}: \#10 (Pagel et al No. 1); \#20 (No. 5)
and \#24 (No. 8). For \#20, the extinction is the
same within the errors between both the long slit data of Pagel et al
\shortcite{PA80} and the fibre spectroscopy, although the
excitation of the fibre spectra is lower (stronger \oii and
weaker \oiii). For the two other regions the extinction differs
considerably (by about 0.3 in c) whilst the excitations are
similar. Detailed comparison of both sets of spectra is not warranted
since the sampling region is different and it is known from previous
work (e.g. \cite{RW87}, \cite{RO89}) that extinction and 
ionization conditions
are not constant across the surface of resolved extragalactic
\hii regions.

\begin{table*}
\caption{H {\sc ii} regions of NGC 1313 -- Reddening-corrected line fluxes (H$\beta$ = 100)}
\begin{tabular}{rrrrcccccccc}
\hline \hline
No& X     &    Y   &    [OII]  &       [OIII]  &       HeI    &   [NII] &    [SII]  &     [Ar III]  &    c   & $\rho/\rho_0$ \\ 
  & $''$     &   $''$  &    3727  &     5007 &      5876    &    6584  &    6717-30 &       7136   &        &       \\
(1) & (2) &  (3) &     (4)   &       (5)       &       (6)     &    (7)  &    (8)     &      (9)   &    (10)  &   (11)\\
\hline
1 & 183   &    21  &   995$\pm$118 &  181$\pm$24 &     &  38$\pm$11 &  115$\pm$17  &      &  0.46$\pm$0.38 &  1.00 \\
2 & 174   &    43  &   485$\pm$49 &  179$\pm$18 &     &  37$\pm$6 &   71$\pm$8  &      &  0.76$\pm$0.28 &  0.96 \\
3 & 158  &     21  &   358$\pm$7 &  323$\pm$5 &  10.6$\pm$0.8&  17$\pm$1 &   38$\pm$2  &   10$\pm$1  &  0.85$\pm$0.05 &  0.87 \\
4 & 153  &    -16 &    573$\pm$11 &  232$\pm$4 &  10.6$\pm$0.9&  19$\pm$1 &   34$\pm$2  &    9$\pm$1  &  0.93$\pm$0.05 &  0.84 \\
5 & 132  &     17 &    340$\pm$3 &  425$\pm$3 &  10.4$\pm$0.3 & 18$\pm$1 &   37$\pm$1  &    9$\pm$1  &  0.96$\pm$0.02  & 0.73 \\
6 & 140  &     64 &    202$\pm$13 &  441$\pm$20 &  14$\pm$3  &  12$\pm$3 &   49$\pm$6  &     &   0.70$\pm$0.15 &  0.80 \\
7 &  85  &     45  &   276$\pm$6 &  326$\pm$5 &   9.7$\pm$0.8 & 14$\pm$1 &   30$\pm$1  &    8$\pm$1 &   0.83$\pm$0.05 &  0.50 \\
8 &  65  &     45  &   281$\pm$7 &  321$\pm$6 &   8.0$\pm$1.2&  14$\pm$1  &  26$\pm$1   &   9$\pm$1 &   0.67$\pm$0.06 &  0.39 \\
9 &  76   &    70  &   436$\pm$7 &  189$\pm$3  &  7.5$\pm$0.7 & 25$\pm$1 &   60$\pm$2  &    7$\pm$1  &  0.98$\pm$0.04 &  0.49 \\
10&  46  &     71  &   301$\pm$21 &  254$\pm$17 &      &  34$\pm$3 &   51$\pm$6  &      &  0.61$\pm$0.20 &  0.37 \\   
11&  51  &    107  &   365$\pm$6 &  280$\pm$4 &  11.0$\pm$1.0 & 23$\pm$1  &  44$\pm$1  &   11$\pm$1 &   0.72$\pm$0.05 &  0.48 \\  
12&  27  &     68 &    230$\pm$2 &  373$\pm$3 &  10.5$\pm$0.3 & 13$\pm$1 &   26$\pm$1  &    9$\pm$1 &   0.72$\pm$0.02 &  0.29 \\
13&  17  &     49  &   360$\pm$7 &  192$\pm$3 &  11.5$\pm$0.9 & 30$\pm$1 &   47$\pm$1  &    7$\pm$1 &   0.85$\pm$0.05 &  0.20 \\
14&  -6 &      42  &   592$\pm$51 &  161$\pm$15 &   &   45 $\pm$5  &  90$\pm$10  &     &  0.92$\pm$0.26 &  0.16 \\
15&  39 &      13  &   771$\pm$50 &  106$\pm$10 &    &   50$\pm$5 &  138$\pm$11  &      &  0.60$\pm$0.20 &  0.22 \\
16&  52  &    -33  &   179$\pm$4 &  416$\pm$7 &    &   14$\pm$1 &   25$\pm$1  &   12$\pm$1 &   0.66$\pm$0.06 &  0.31 \\
17&  -2  &    -27  &   571$\pm$29 &  197$\pm$10 &    &   49$\pm$4 &   87$\pm$7  &     &   0.66$\pm$0.15 &  0.10 \\
18&  -9 &     -48  &   602$\pm$12 &  121$\pm$3 &   9.0$\pm$1.4 & 38$\pm$1&    77$\pm$2 &     8$\pm$1 &   0.64$\pm$0.06  & 0.18 \\
19&  14   &   -77 &    336$\pm$10 &  160$\pm$5 &   7.1$\pm$1.8&  29$\pm$1 &   46$\pm$2 &     9$\pm$2  &  0.67$\pm$0.08 &  0.30 \\
20& -54  &    -31 &    395$\pm$13 &  210$\pm$6 &   7.4$\pm$1.5 & 31$\pm$2 &   78$\pm$3 &     9$\pm$1  &  1.04$\pm$0.09 &  0.32 \\
21& -74 &     -22 &    281$\pm$3 &  340$\pm$2 &   9.9$\pm$0.3 & 17$\pm$1 &   31$\pm$1 &    10$\pm$3 &   0.94$\pm$0.02 &  0.41 \\
22& -93 &     -13 &    427$\pm$8 &  281$\pm$5 &  11.2$\pm$0.8 & 24$\pm$1 &   51$\pm$1 &    10$\pm$1&    1.18$\pm$0.05 &  0.51 \\
23& -46  &    -87  &   155$\pm$2 &  474$\pm$4 &  12.0$\pm$0.5 & 15$\pm$1 &   31$\pm$1 &    11$\pm$1 &   0.51$\pm$0.03 &  0.41 \\
24& -215 &     -98  &   439$\pm$21 &  205$\pm$10 &     &  22$\pm$3 &   71$\pm$6 &      &   0.77$\pm$0.14 &  1.23 \\
25& -57  &   -162  &   609$\pm$71 &  271$\pm$29 &     &  45$\pm$7 &   52$\pm$10 &       &  1.06$\pm$0.32 &  0.67 \\
26& -87   &  -169 &    404$\pm$10 &  191$\pm$4 &   7.6$\pm$1.5&  19$\pm$1 &   44$\pm$2 &     8$\pm$1  &  0.82$\pm$0.07 &  0.78 \\ 
27& -102  &   -195 &    529$\pm$9&   176$\pm$3 &  10.0$\pm$1.0 & 23$\pm$1 &   55$\pm$1  &    9$\pm$1 &   0.61$\pm$0.05 &  0.91 \\
28&  49   &  -185  &   596$\pm$20 &  317$\pm$10 &   9.0$\pm$2.1 & 22$\pm$2  &  38$\pm$2 &    11$\pm$2 &   0.75$\pm$0.10 &  0.73 \\ 
29& -282 &      23 &    336$\pm$15 &  316$\pm$12 &   9.5$\pm$2.6 &  17$\pm$2 &   43$\pm$4  &      &  0.67$\pm$0.12 &  1.54 \\
30& -246  &    -79  &   755$\pm$142 &  196$\pm$39  &    &     &      &     &  0.81$\pm$0.58 &  1.37 \\
31& -183  &    113  &   229$\pm$13 &  263$\pm$13 &   &  17$\pm$3  &  49$\pm$6    &    &  0.55$\pm$0.15 & 1.08 \\
32& -126  &   -295  &   325$\pm$11  & 202$\pm$7 &  11.2$\pm$1.7& 21$\pm$2  &  48$\pm$2  &      &  0.78$\pm$0.10 & 1.28 \\
33&  -68  &    -49  &   996$\pm$175  & 188$\pm$35  &   &  38$\pm$9 &  132$\pm$20  &      &  1.23$\pm$0.53 & 0.42 \\
ISM1&  -21  &      0  &   710$\pm$70 &  101$\pm$14 &    &  40$\pm$7 &  132$\pm$10  &      &  0.94$\pm$0.30 & 0.12 \\
ISM2&  41   &    39  &   710$\pm$154 &  103$\pm$28 &    &  73$\pm$19 &  171$\pm$30  &      &  0.64$\pm$0.67 & 0.27 \\
ISM3&  -80  &   -197  &   179$\pm$29  & 186$\pm$26  &   &  43$\pm$10 &   95$\pm$18  &       & 0.59$\pm$0.42 & 0.85 \\
\hline
\end{tabular}
\end{table*}

Various diagnostic diagrams are plotted in Figure 3 against the sequencing index
\oiioiii. The spread in values is small attesting to the rather high excitation and
small range in ionization conditions. Both the diagnostic diagrams \niioii and \siioii versus \oiioiii
(Fig. 3) are characteristic of high excitation nebular
gas. From the point of view of physical
conditions, the \hii regions of NGC 1313 are
similar to the highest excitation \hii regions
found in the outer parts of the spiral
galaxy M 101 (see Kennicutt \& Garnett \shortcite{KG96}); in contrast NGC
2997 does not appear to possess this high excitation
population (see Walsh \& Roy \shortcite{WR89}).
The \hii regions of NGC 1313 illustrate the narrow range of thermal conditions
and the crowding effect close to log \oiioiii $\sim$ 1.0
as discussed by  McGaugh \shortcite{MG91}. The optical lines of
oxygen dominate the cooling and their strengths are driven mainly
by the heating rate. However the values of \ohb and \oiioiii
might be in the degenerate range where they become nearly 
independent of the level of O/H (McGaugh \shortcite{MG91}; Roy et al.
\shortcite{RO96A}). Then the strength of \oiihb and \ohb depends
mostly on the effective temperature of the ionizing radiation field
and on the ionization parameter.

\begin{figure*}
\centerline{\psfig{file=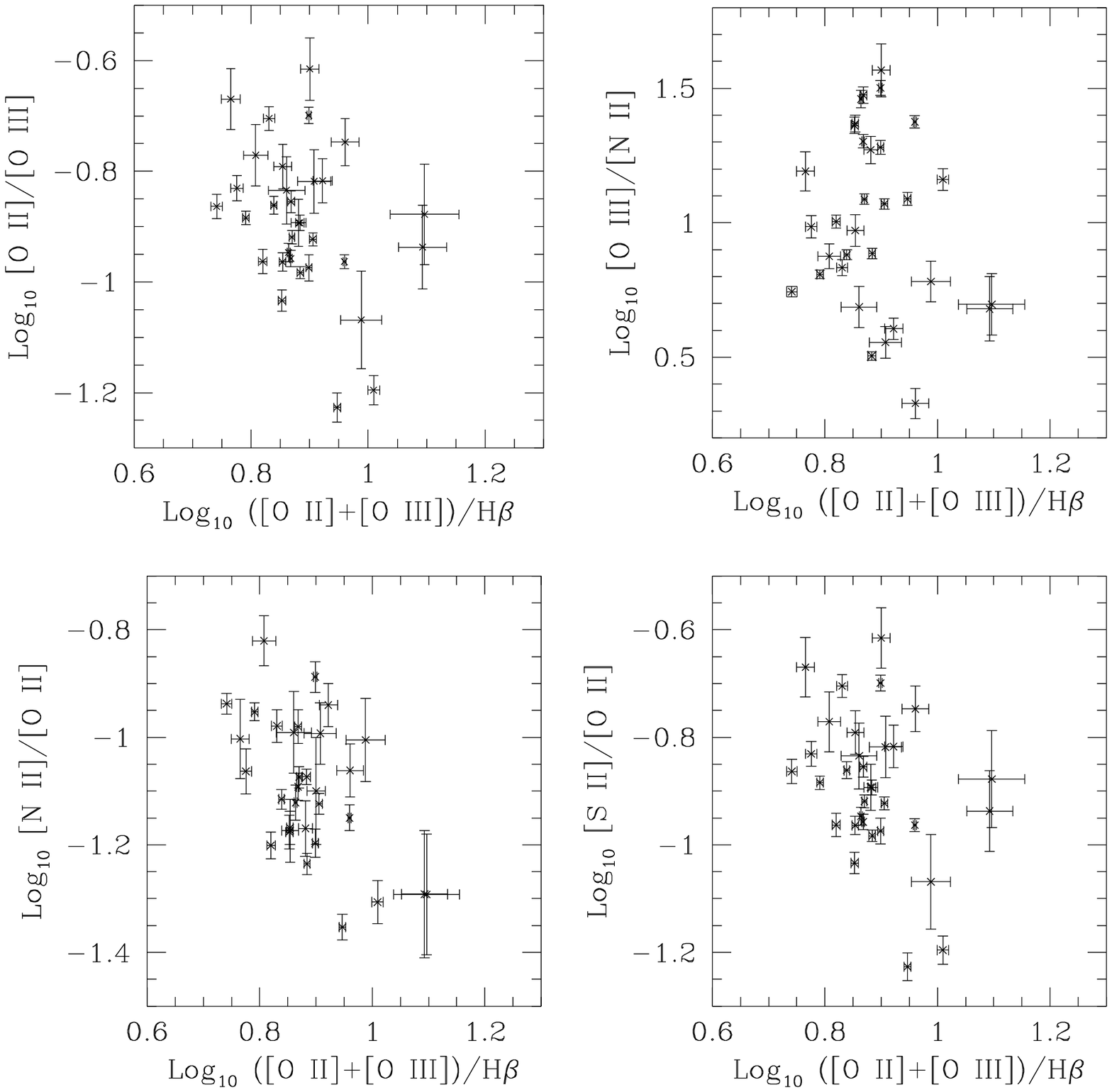,height=16.0cm,clip=}}
\caption{Diagnostic diagrams of log \OiiOiii, log \oiiinii, log \niioii 
and log \siioii , vs. the sequencing index log \oiioiii for the 33 \hii regions
in NGC 1313.}
\label{fig3}
\end{figure*}

 Although no obvious trend is seen across the disc of NGC 1313,
the radial behaviour of some of the nebular parameters is illustrated in
Figure 4. The radial position
of the \hii regions are given in Table 2 (column 11) in terms of
the fractional isophotal radius (Table 1).
Figure 4 shows $c$, the logarithmic extinction at \Hb, \oiioiii and \oiiinii 
versus radial
distance in terms of the fractional isophotal radius $\rho_0$ (RC3).
The behavior of these nebular indicators is rather uniform
across the disc of NGC 1313. The exception
is the \hii regions of the bar which show slightly lower values of the
\oiiinii ratios.

The three ISM positions show relatively high ratios of \nha and
\sha (Table 2). This is especially true of \sha which has a mean
of 0.46 $\pm$ 0.13 in the ISM locations compared to 0.20 $\pm$ 0.10 in
the \hii regions. The values measured in the ISM of NGC 1313 are very
similar to those found by Rand \& Kulkarni \shortcite{RK90}
in the extraplanar ionized gas of NGC 891, or by Reynolds \shortcite{RE88}
in the diffuse interstellar background of the Galaxy. Rand \& Kulkarni
have suggested a diffuse radiation field from a Population I source as
the origin of the warm diffuse ionized gas.

\begin{figure*}
\centerline{\psfig{file=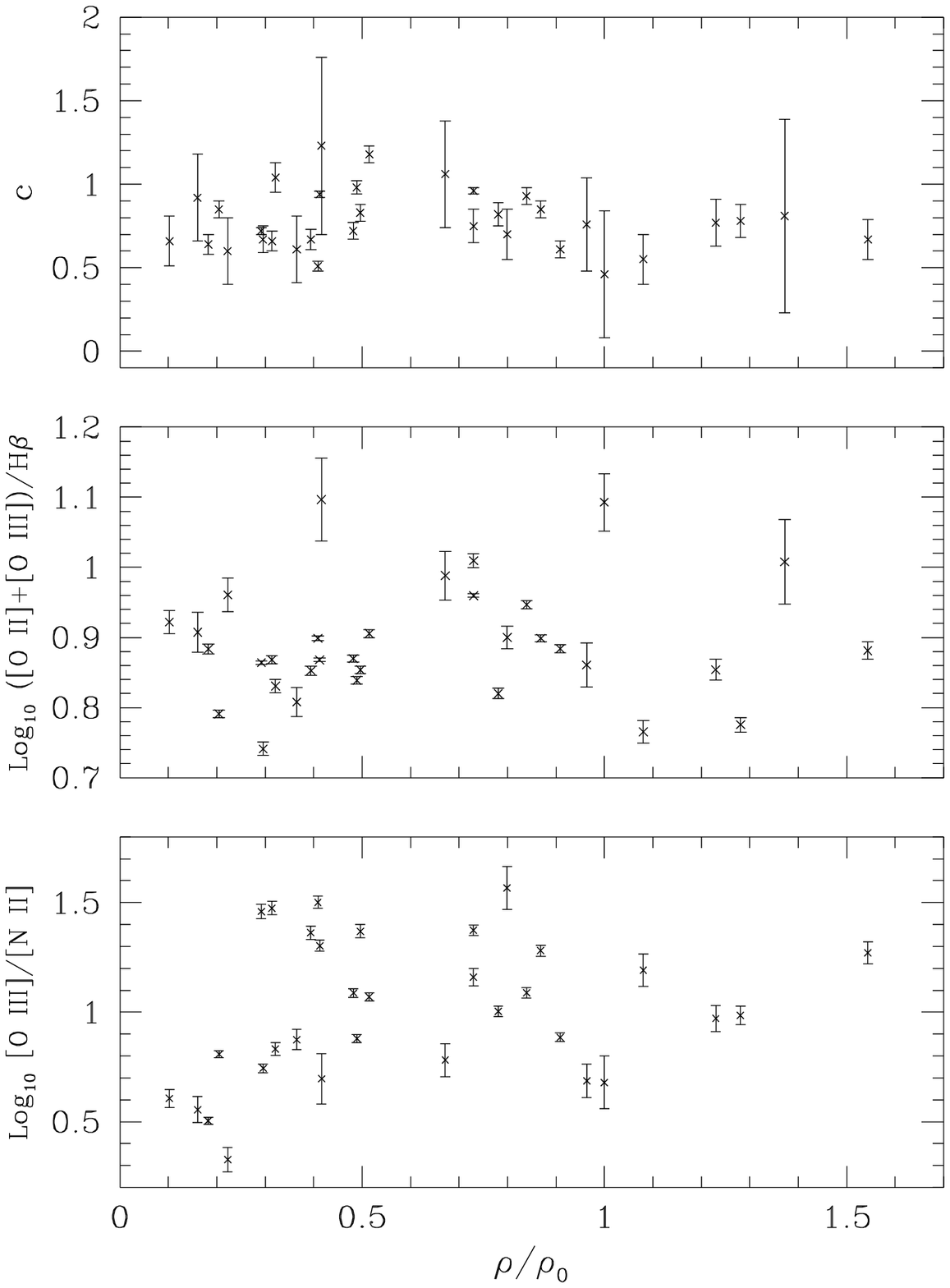,height=14.0cm,clip=}}
\caption{Various observed parameters of the NGC 1313 \hii regions are 
plotted against the normalized isophotal radius:
extinction  $c$(\Hb) (top); log \oiioiii (middle); and log \oiiinii (bottom).}
\label{fig4}
\end{figure*}

\section{Discussion}

\subsection{The O/H abundance distribution}

Using the \oiioiii ratio to derive the  O/H abundance
distribution in NGC 1313 (Edmunds \& Pagel \shortcite{EP84},
McCall et al. \shortcite{MC85}, Dopita \& Evans \shortcite{DE86}), one obtains
a flat radial distribution at 12 + log O/H = 8.33 $\pm$ 0.14.
It seems preferable to use the ratio \oiiinii
to derive O/H abundances since \oiioiii could be in the
degenerate regime with respect to oxygen abundance (McGaugh \shortcite{MG91};
Roy et al. \shortcite{RO96A}).
The calibration of \oiiinii by Edmunds \& Pagel \shortcite{EP84} was used
to derive O/H with the following polynomial fitting for the range appropriate
to the low O/H abundances and small values of \oiiinii :

\begin{equation}
12 + {\rm log~O/H} = 8.77 - 0.214x - 0.407 x^2 + 0.406 x^3 - 0.118 x^4,
\end{equation}
\noindent where $x =$ log \oiiinii $>$ --0.18 \cite{RO96A}. The abundance distributions, 
 shown in  Figure 5 are consistent with a flat gradient
(mean 12 + log O/H = 8.44 $\pm$ 0.09 and slope $-0.25~\pm$ 0.16 dex/$\rho_0$)
for most of the body of the galaxy (r $\geq$ 100$''$).
When using \oiiinii as an abundance indicator, one obtains O/H values 
higher by more than 0.1 dex in the inner \hii regions
corresponding to the bar; the bar extends to about $r = 70''$ or
$\rho$ = 0.25 $\rho_0$ (1.5 kpc).

Abundances could be directly measured
in four \hii regions using the electron temperature determined 
from the \oiii 4363\AA\ line; collision strengths and transition probabilities 
were taken from Mendoza \shortcite{ME83}. These points are shown by filled circles
on Figure 5. The discrepancy between the directly determined O/H
abundance and that from the strong line calibration is indicated by a bold line
for the four \hii regions on Figure 5. The mean temperature
of the four (11,000K) was also applied to the 29 \hii region spectra without 
measured T$_e$ to redetermine the O/H abundances 
using the strength of the \oii and \oiii lines.
The resulting abundance distribution is again flat at 12 + log O/H = 8.32
$\pm$ 0.08, with slope -0.05 $\pm$ 0.15 dex/$\rho_0$. The abundance distribution is uniform, with perhaps a slight increase in the bar.
The bar is not as prominent in \Ha\ as the arms are. However the possibly higher
level of O/H suggested by the \oiiinii ratios in the bar
is suggestive of recent enrichment which has not yet been diluted by the radial flows.
In summary, whichever calibration is used, the O/H distribution
across NGC 1313 appears uniform, as already implied in the earlier work of
Pagel et al. \shortcite{PA80} using a much smaller number of \hii
regions.

\begin{figure*}
\centerline{\psfig{file=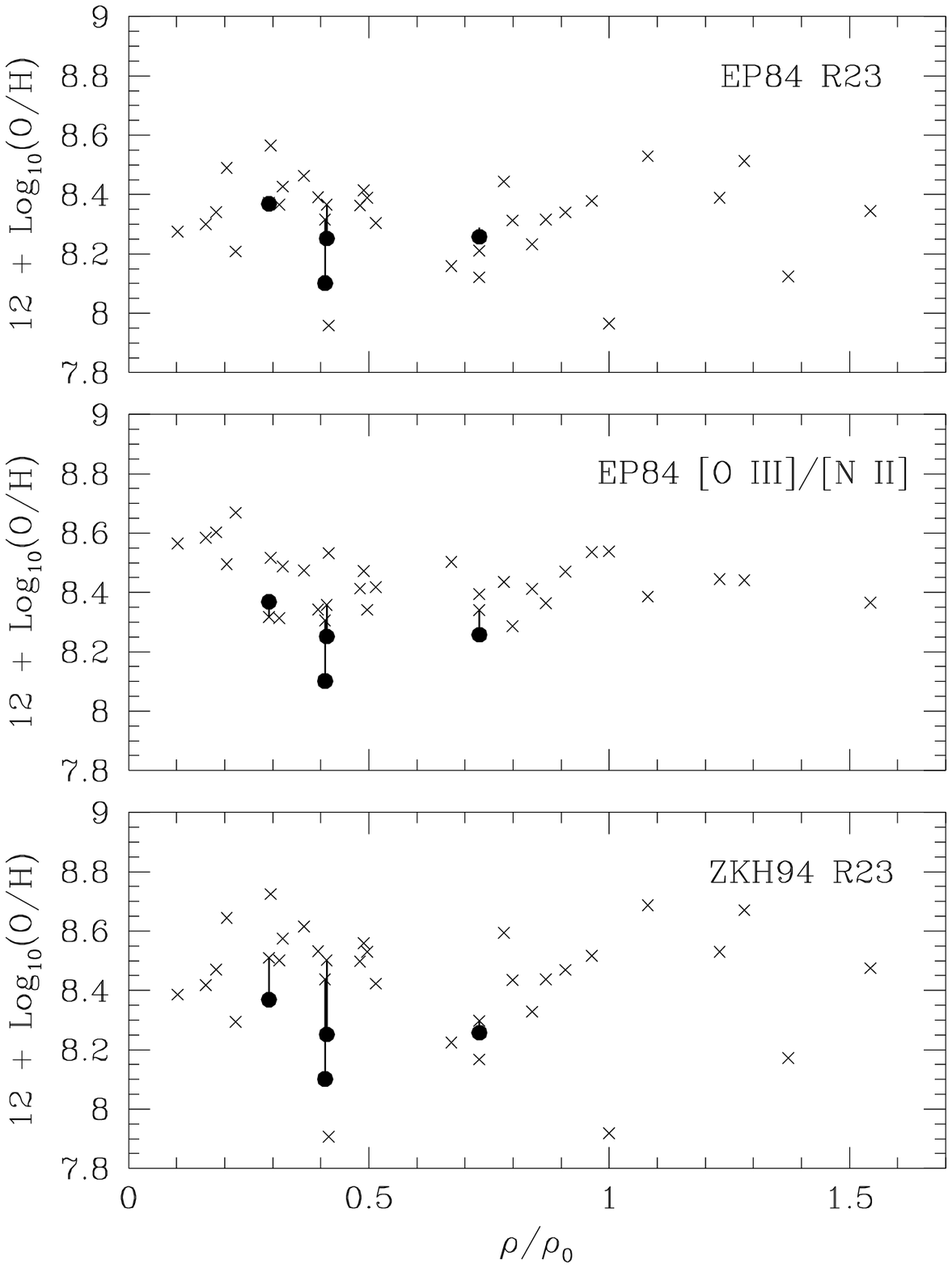,height=16.0cm,clip=}}
\caption{(top) The radial oxygen abundance distribution across NGC 1313 
using the calibration of \oiioiii by Edmunds \& Pagel (1984).
The same distribution using the calibration
of \oiiinii by Edmunds \& Pagel (1984) (middle) and Zaritsky et al 
(1994) (bottom). The black circles
are the derived abundances based on the \oiii electron
temperatures for the regions where the \oiii 4363\AA\ line
could be measured. The bar extends to $r \sim 70''$ or about $\rho/\rho_0$
= 0.25.}
\label{fig5}
\end{figure*}

\begin{table*}
\caption{Properties of some small disc galaxies} 
\begin{tabular}{lcrrc}
\hline \hline
Object & Type & $M_B$ & V$_{\rm max}$ & O/H gradient\\
      &  &  (mag.)   & km/s & (dex/kpc)\\
\hline
SMC & IM IV-V & -16.99&  & 0.00\\
LMC & SBm III & -18.43 & 65 & 0.00 \\
NGC 2366 & SBm IV-V & -16.73& 46 &-0.003$\pm$0.015\\ 
NGC 4214 & SBm III & -18.79& 59 & 0.00 \\
NGC 4395 & Sd III-IV & -18.57& 90 & 0.013$\pm$0.016\\
NGC 1313 & SB(s) III-IV & -19.66& 116 & $<$-0.04$\pm$0.03\\
NGC 300 & Sc II.8 & -18.59& 102 & -0.18\\
NGC 598 & Sc(s) II-III & -19.07& 107 & -0.13\\
NGC 7793 & Sd(s) IV & -18.85& 116 & -0.16\\ 
\hline 
\end{tabular}\\
Notes: (a) M$_{\rm B}$ and morphological types
 are from Sandage \& Tammann \shortcite{ST81}. 
(b) V$_{\rm max}$ are from Luks \& Rohlfs \shortcite{LR92} (LMC),
from Wevers (thesis) (NGC 2366, NGC 4395), Allsopp \shortcite{AL79}
(NGC 4214), 
Puche et al. \shortcite{PU90} (NGC 300), Ryder et al. \shortcite{RY95}
 (NGC 1313),
Rogstad et al. 1976 \shortcite{RO76} (NGC 598), and Carignan \& Puche (1990) (NGC 7793). (c) Abundances data are from
 Pagel et al. \shortcite{PA78} (LMC), Kobulnicky \& Skillman \shortcite
{KS96} (NGC 4214);
 Roy et al. \shortcite{RO96A} (NGC 2366, NGC 4395); Zaritsky et al.
\shortcite{ZA94} (NGC 300,
NGC 598, NGC 7793).
\end{table*}

\subsection{Behaviour of abundance gradients in disc galaxies}

In Table 3 are presented brief data on other low luminosity disc galaxies.
The absolute B luminosity is given in column 3 and the 
O/H abundance gradient in column 4. NGC 300 and NGC 7793 are the 
smallest known galaxies having a well-developed steep negative
O/H radial abundance gradient.
Inspection of Table 3 reveals other galaxies as luminous as these two,
which have no gradient. In particular NGC 1313 seems at least as luminous 
and massive as
M 33 (NGC 598) and NGC 7793, but it does not have a global O/H
gradient. NGC 1313 differs from NGC 598 and NGC 7793 mainly by its bar feature. 

Bars can modify the chemical evolution
of disc galaxies as shown by Friedli \& Benz 
\shortcite{FB95} and Friedli et al. \shortcite{FR94}. Bar-driven
gas fueling the central regions results in enhanced star
formation and metal-enrichment. But as the bar ages and
re-fueling by the bar dies out, large scale dilution
by radial mixing will decrease the slope
of any pre-existing radial gaseous or stellar abundance
gradient. A galaxy with an old and strong  bar should display
a shallow gradient. The bar of NGC 1313
is gas-rich; star formation may have been vigorous 
if the apparent metal enhancement (given only by the \oiiinii 
indicator) is real. Thus its bar would have to be young. Young and strong
bars which are gas rich and forming massive stars should
have a steep abundance gradient at least
in the bar (Friedli et al. \shortcite{FR94}), but NGC 1313 does not.
This may be related to the moderate strength of its bar whose stellar 
axis ratio $b/a(i)$ = 0.63 (Martin \shortcite{MA95}).

The character of the radial abundance distribution in SBm and
late SBc galaxies obviously requires the effect of bars to be re-examined
when applied to smaller mass disc galaxies.
Clearly these galaxies have either a strong global gradient
or no gradient at all, as exemplified by Table 3. The subtle effect of bars in galaxies
of larger mass, which is to produce a shallow gradient, becomes
in small mass disc galaxies a purely ``on'' or ``off'' effect.
As soon as a bar is present in a small mass galaxy, the global abundance distribution
-- at least as seen in the interstellar gas -- becomes uniform.
This implies that the homogenizing action of the bar is
rapid, or that systems like the Magellanic Clouds or NGC 1313 always had a bar
or, less likely, never had an abundance gradient. The
presence of a bar in low-mass disc galaxies  results in the
absence of any abundance gradient.

Although no significant correlation appears to exist between the
slope of abundance gradients, the rotational velocity and the
morphological type of the galaxies \cite{ZA94},
only late-type systems are able to have either a steep single-scale
exponential radial abundance profile or no gradient at all. Moving
to earlier types, gradients are observed in all galaxies,
but they are never as steep as in late types. Moreover, if
they have a young bar like NGC 3359 \cite{MR95},
NGC 3319 \cite{ZA94} and NGC 1365 
\cite{RW96}, the gradients show a break, with a steep
inner part and a flat abundance distribution in the outer parts.

For axisymmetric discs, models show only a narrow range
of gradient slopes and these evolve little with time (e.g.
Moll\'a et al. \shortcite{MO96}). Such models  apply obviously
only to a narrow range of ``normal'' galaxies. There is need
for theoretical investigation of the more common type of galaxy discs.
Two thirds of all galaxies have a bar or a non-axisymmetric
oval stellar light distribution in their centre (Sellwood \& Wilkinson
\shortcite{SW93}; Martin \shortcite{MA95}). These asymmetries induce large scale radial
flows and mixing, and the galaxy must evolve very differently
from those where the different radial sections are considered
as closed boxes. Dramatic effects
on the shape and the slope of the radial abundance
distribution can appear. These effects may take place over
a period of a few dynamical timescales only;
these timescales may be short, as the ``on'' or ``off'' effect of bars on
abundance gradients in late-type galaxie would suggest. Mechanisms capable of
producing such a uniform abundance distribution have been
explored by G\"otz \& K\"oppen \shortcite{GK92}, Edmunds \& Greenhow 
\shortcite{EG95}  and Roy et al. \shortcite{RO96A}, but no detailed 
hydrodynamical simulations have been performed corresponding to evolving small
gas-rich galaxies.
\\

\noindent
{\bf Acknowledgments} \\
We thank J. Pogson for fine technical support at the AAT. 
We acknowledge helpful discussions with Danield Friedli, Pierre Martin, 
St\'ephanie C\^ot\'e and Laurent Drissen. 
The PATT Committee are thanked for assigning time on the AAT for this
project. This investigation was funded in part by the Natural
Sciences and Engineering Research Council of Canada, by the Fonds
FCAR of the Government of Quebec and by
the Visitor Program of the European Southern Observatory.

\end{document}